# Effective Requirements Generation: Synchronizing Early Verification & Validation, Methods and Method Selection Criteria

Lester O. Lobo and James D. Arthur

Department of Computer Science, Virginia Tech

Blacksburg, Virginia 24060, USA

{lester, arthur}@vt.edu

**ABSTRACT.** This paper presents a framework that guides the requirements engineer in the implementation and execution of an effective requirements generation process. We achieve this goal by providing a well-defined requirements engineering model that includes verification and validation (V&V), and analysis. In addition, we identify focused activity objectives and map popular methods to lower-level activities, and define a criterion based process for optimizing method selection for attendant activities. Our model, unlike other models, addresses the complete requirements generation process and consists of activities defined at more adequate levels of abstraction. Furthermore, our model also incorporates a unique approach to V&V that enhances quality and reduces the cost of generating requirements. Additionally, activity objectives are identified and explicitly stated - not implied as in the current models. Activity objectives are crucial because they drive the selection of methods for each activity. To assist in the selection of an appropriate set of methods, we have mapped commonly used methods to activities based on their objectives. Finally, we have identified method selection criteria and prescribed a reduced set of methods that optimize these criteria for each activity in our model. Thus, our approach assists in the task of selecting methods by using selection criteria to reduce a large collection of potential methods to a smaller, manageable set. The model, clear mapping of methods to activity objectives, and the criteria based process, taken together, provide the much needed guidance for the effective implementation and execution of the requirements generation process.

# Effective Requirements Generation: Synchronizing Early Verification & Validation, Methods and Method Selection Criteria

## 1. INTRODUCTION

The objective of software engineering is to develop and adapt software systems to satisfy user's needs, schedule and budget constraints. In pursuit of this goal, a substantial amount of research has been conducted in improving the software development process. However, according to the Standish report, only 28% of the real world projects are successful [1]. Williams attributes this low rate of success primarily to the lack of clear and precise requirements [2]; the reason being that a system is only as good as the requirements from which it is developed. This finding indicates that the industry still lacks an effective definition of the requirements generation process. The prevailing uncertainty is because of shortcomings in three different areas – adequacy of the requirements engineering model, identification of activity objectives and their synchronization with methods, and guidance in selecting methods for activities within the model.

The first problem area is derived from the observation that many models do not adequately address the requirements generation process. Current requirements engineering models such as the Win-Win model [3], Requirements Triage [4], and RE Process Framework [5] either have a narrow focus on only portions of the requirements generation process, or provide a broader perspective defined by abstract, high-level activities. As a result, the requirements engineer lacks a definitive model that guides him/her through the requirements generation process.

Another problem area is that current models often include implicit activity objectives. Because these models have an inadequate level of abstraction, they lack a clear mapping of objectives to activities. Activity objectives are critical because they drive the selection of methods/techniques for the activities in the model. It is widely acknowledged that methods have significant impact on the quality of the final product [6]. Hence, a substantial amount of research has been conducted in identifying methods for the entire software development life cycle. As a result, there are a large number of methods for the requirements engineering process [7] [8] [9] [10] [11]. To date, however, these methods are mapped to the high-level, abstract activities (e.g. elicitation, analysis, specification); there is a noticeable lack of the coordination of methods with lower level activities [12]. Thus, without explicit objectives and mapping between methods and objectives, it is likely that the requirements engineer will overlook certain objectives and its associated methods; this can have an adverse impact on the projects success.

The third problem area that a requirements engineer faces is the lack of guidance in selecting appropriate methods for the activities in the requirements engineering model. Even if we do synchronize methods to lower level activities/objectives, the collection of methods to choose from is fairly large. There is no guidance when a particular method should be chosen over the other. Given this scenario, the requirements engineer often selects methods in an *ad-hoc* fashion, resulting in an output which insufficiently addresses the objectives of activities in the requirements model.

This paper describes a framework that's helps the requirements engineer in overcoming the three problem areas discussed above. We propose a two-phase requirements engineering model that is well-defined and which addresses the entire requirements generation process. In addition, the model consists of activities decomposed at an adequate level of granularity to facilitate the selection of methods. An added advantage of the model is that it includes a unique approach to verification and validation (V&V) that enhances the quality of the requirements generated, and reduces the time and effort associated with the overall V&V activities. Because activity objectives drive the selection of methods, the objectives of each activity in the model are identified and explicitly stated, that is, there are no implied objectives. Additionally, we have identified methods commonly used in the industry and mapped them to the decomposed activities, based on their stated objectives. Consequently, the requirements engineer has an appropriate set of methods to choose from for each activity objective. In order to simplify the task of selecting methods from a collection of suitable methods, we have grouped these methods according to criteria they purport to optimize, e.g., cost, personnel, time, or completeness. Thus, given a selection criteria, the requirements engineer can easily trace a path of methods (within the reduced set) that optimize the chosen criteria for the entire requirements generation process.

The remainder of the paper is structured as follows: In Section 2, we discuss the proposed two-phase requirements engineering models and identifies the benefits derived from the model. In Section 3, we describe the identification of activity objectives and their synchronization with the appropriate set of methods. Section 4 identifies the method selection criteria and discusses the process of choosing the sequence of methods to optimize the chosen selection criteria. Finally, Section 5 presents the summary and possible future work.

## 2. THE PROCESS MODEL

The requirements engineering phase typically consists of elicitation, analysis, specification and verification activities [13]. In this paper, we address these activities with the goal of generating a clear and complete set of requirements in a cost effective manner. Our proposed model focuses on early V&V to alleviate several problems associated with software development [14][15]. The effect is that the cost incurred during product development is minimized through early error detection and correction. Another goal of the model is to represent the linking of requirements engineering activities at a level of decomposition that closely reflects activities in the requirements engineering process practiced in industry today. Thus, our model addresses the inadequacies of many of the earlier approaches that address the requirements engineering phase at more abstract levels, and thus making those models more difficult to understand and implement.

Briefly, our approach consists of the following two phases:

- **Local analysis**: an iterative phase concentrating on eliciting, analyzing, documenting, and evaluating small sets of related requirements.

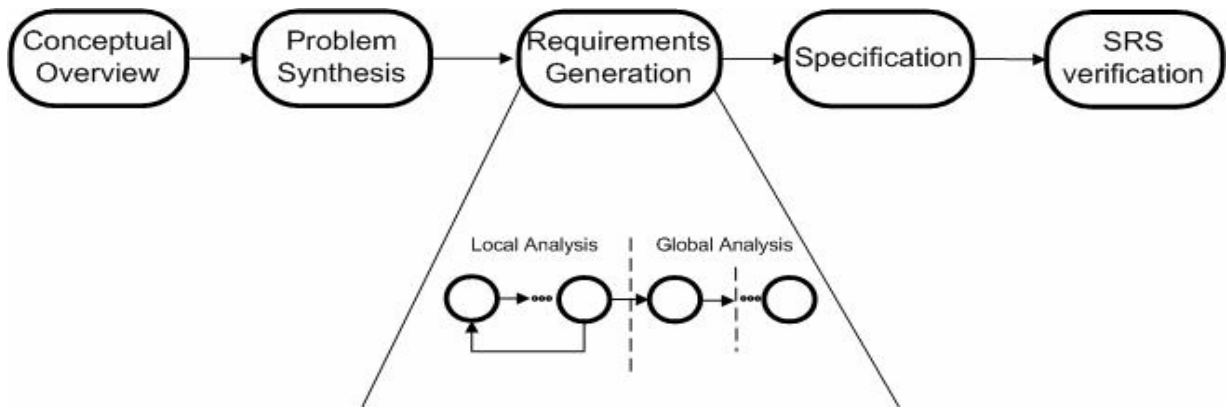

**Figure 1. The Local and Global Analysis phases in the context of the RGM**

- **Global analysis**: a set of activities that complement the Local Analysis phase and focus on selected verification and related business concerns of the more comprehensive set of requirements.

A brief description of the RGM is provided in the subsequent section.

## 2.1 The RGM

Our modeling approach extends the Requirements Generation Model (RGM) [16] which addresses the complete requirements engineering phase from a relatively high-level perspective. The RGM consists of five phases shown in Figure 1.

The first two phases represent the problem domain; they gather information about the existing problems and needs. The remaining phases correspond to the solution domain; they capture the requirements of the solution. The first phase, Conceptual Overview, helps in recognizing the need for the new system both from the business and operational perspectives. The Problem Synthesis phase assists in identifying the customer problems and needs. The Requirements Generation phase helps elicit, analyze and produce a complete set of requirements which adhere to established quality characteristics. In the Specification phase these requirements are then incorporated into a formal SRS. Finally, the SRS is verified and baselined before submitting it to the customer for approval.

Our focus is on the critical Requirements Generation phase because this phase transforms the customer needs into concrete software requirements. We have decomposed the Requirements Generation phase into activities based on the concept of "Separation of Concerns" which focuses on identifying and satisfying a small set of concerns for organizing and decomposing complex processes [17][18]. Our primary concerns during the decomposition of the model were that: (1) activities should have a focused objective and (2) the decomposition into activities should facilitate the selection of methods for identified activities based on their objectives. By using the RGM as an extendible basis, we contend that our model integrates seamlessly into the requirements engineering life cycle.

The next two sections provide a detailed explanation of the expansion of the RGM consisting of the Local and Global Analysis phases.

## 2.2 Local Analysis Phase

Local Analysis is the initial component of the expanded RGM. It is an iterative process through which the customer and the requirements engineer discover, review, articulate and evaluate the sets of requirements for the proposed software system. Entering into the Local Analysis phase, we have the Needs Document, which records the customer problems and needs. The decomposition of the iterative Local Analysis phase into its constituent activities is illustrated in Figure 2.

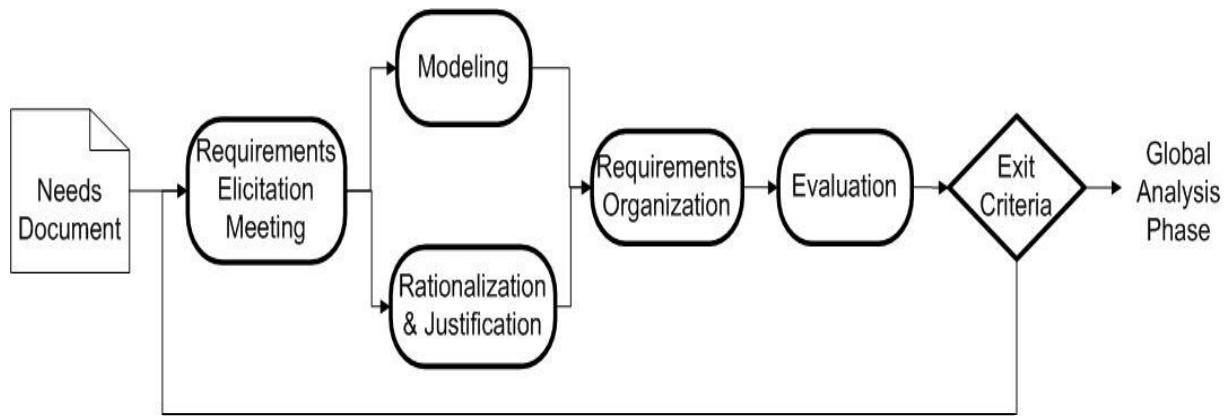

**Figure 2. Local Analysis phase and its activities**

The Needs Document, generated during the Problem Synthesis phase, is the input to the first activity – Requirements Elicitation Meeting. The objective of the elicitation meeting is to correctly identify and capture requirements of the stakeholders. The roles of the requirements engineer and the stakeholder for this activity are complementary - one conveys all necessary system information, the other elicits and captures the requirements and their context. Several approaches such as Joint Application Design (JAD) [19][20], Participatory Design (PD) [21] and Facilitated Application Specification Techniques (FAST) [22] have been effectively utilized for this activity. Some commonly used techniques for elicitation are interviews, brainstorming, and focus groups.

To assist in the analysis of the elicited requirements¸ we introduce the next two activities - Rationalization and Justification, and Modeling. These activities can be executed either in sequence or in parallel. It is well known that these activities are critical for supporting the analysis of requirements. However, most of the current models that address requirements generation, e.g. Knowledge Level Process Model [23] and Win-Win model [3], focus on providing a high-level perspective of the requirements process. As a result, lower level activities such as Modeling are often excluded, obscured, or ignored.

During requirements elicitation, the stakeholders are often vague in the description of their requirements. Hence, we have incorporated the Rationalization and Justification activity to help identify the reasoning behind the requirements. If additional requirements are found embedded within the rationale, their relevance and importance are assessed through additional interaction with the stakeholders. This approach continues to foster collective ownership of documented requirements. The Rationalization and Justification activity also enables tracing the requirements back to the needs through an examination of the requirements rationale/reasoning.

We introduce the Modeling activity to help develop a better understanding of the requirements, and to represent them in a clear and comprehensible manner [24]. In addition, the models provide an effective representation for validating requirements with the stakeholders. Furthermore, they are valuable during the Global Analysis and Design phases because those models provide a more precise view of the requirements, their dependencies and their interactions. Hence, the Modeling activity assists in providing a better understanding of the requirements, supports validation, assists in the impact assessment of business concerns, and aids in design of the software system. Various graphical representations, such as Data Flow Diagrams (DFD) [25], Entity-Relationship Diagrams (ERD) [26], and Unified Markup Language (UML) [27] diagrams are used in the industry for modeling purposes. These diagrams assist the understanding and formulation of requirements, and are often included in the SRS as supplementary material.

As a product of Modeling, and Rationalization and Justification activities, the requirements are represented as unordered lists. Hence, we have incorporated the Requirements Organization activity to identify important requirements attributes and to structure the requirements for better understanding and analysis. Some of the major requirements attributes are: associated risk factors, effort needed, importance to the user, and value addition to the final product. The organization of requirements involves hierarchically classifying the requirements on a functional and non-functional basis. The most common techniques used for such classification are affinity analysis [28] and hierarchical decomposition [29].

Evaluation is the final and most crucial activity for achieving our objectives and in differentiating the Local Analysis approach from other model formulations. This activity includes verification, validation, and conflict resolution of requirements. Verification at this stage is conducted on the sets of requirements corresponding to individually distinct functions as illustrated in Figure 3(a). The objective of verification is to evaluate the requirements for quality characteristics such as non-ambiguity, preciseness, verifiability, and the like. Quality attributes such as completeness, traceability, and consistency that mandate the availability of the complete set of requirements, are only partially evaluated during this particular verification activity. Thus, a large part of requirements verification is achieved during Local Analysis, and thereby reducing the effort of verification in the subsequent Global Analysis phase, which focuses primarily on linkages between requirements sets corresponding to functional partitions of the system (Figure 3(b)). Requirements verification is often accomplished through techniques such as inspections [30], audits [31], and reviews [32].

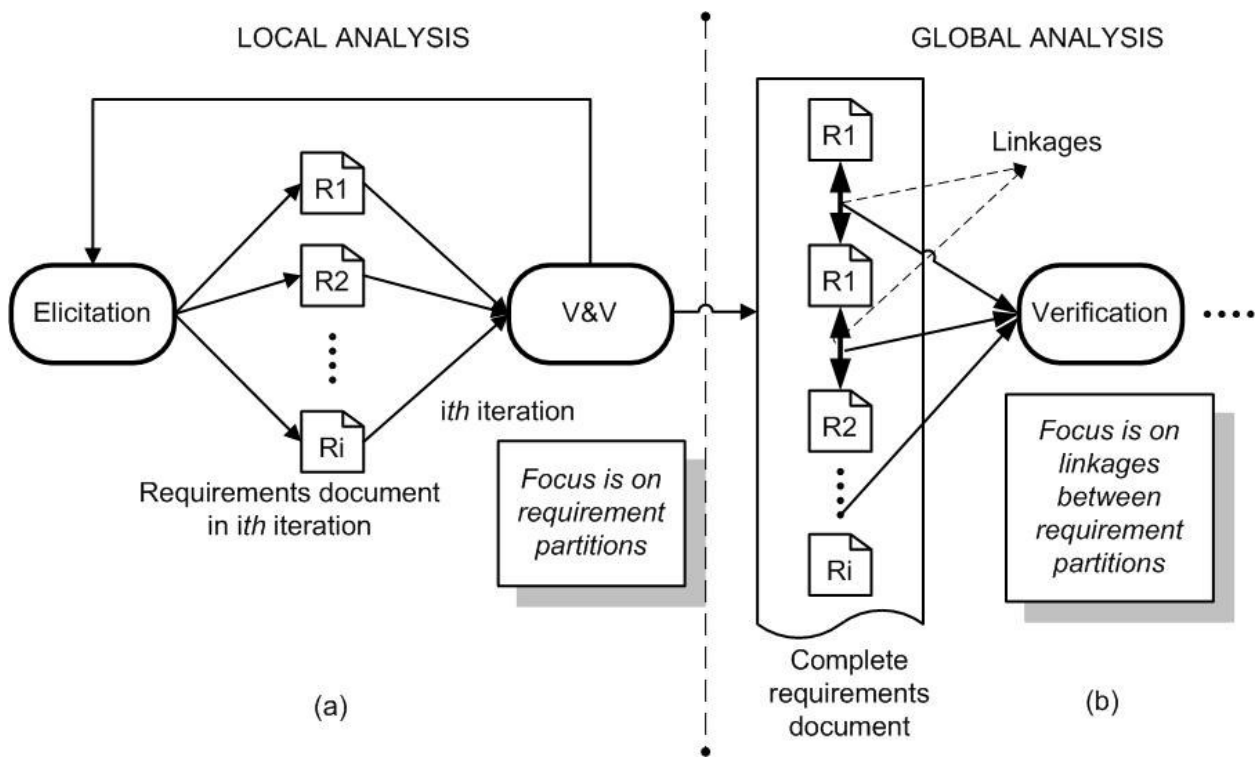

**Figure 3. V&V focus in Local and Global Analysis phases**

The validation component of the Evaluation activity helps determine whether the requirements satisfy the customer's intent. This activity is focused on smaller sets of related requirements - not the complete set (shown in Figure 3(a)). As a result, the stakeholder is more focused on the activity objective, which in turn yields better feedback.

Thus, by performing V&V early in the requirements life cycle we reduce the propagation of requirements errors, enhance the quality of the requirements generated, and thereby reduce the costs incurred during development. Early V&V also takes advantage of the "nearness in time"[1] factor which helps the stakeholder maintain a more focused vision on the recently elicited requirements, and results in better feedback and higher quality requirements. This is in contrast with earlier approaches where V&V is performed towards the end of the requirements phase – after the generation of the formal SRS document. In these approaches, due to the large time gap between V&V and elicitation activities, the requirements details are often obscured in the stakeholders' minds - this has a negative impact on the feedback provided to the requirements engineer. Our model overcomes this drawback by performing V&V on incremental sets of requirements close in time to their elicitation. Requirements inconsistencies identified during V&V are addressed in the conflict resolution activity, which is most effective when the interest based bargaining approach [33] is employed.

---

[1] It refers to the closeness in time of the elicitation and V&V activities that provides the benefits of the clarity of information in the stakeholders' minds.

On completion of the Evaluation activity, it is necessary to determine if another iteration of the Local Analysis phase is needed. This decision is based on exit criteria consisting of a checklist of items that pertain to the following: (a) inspecting requirements quality attributes, (b) ensuring the requirements necessarily and sufficiently trace back to the needs and (c) finding agreement among stakeholders that all requirements have been elicited. The exit criteria listed here are not comprehensive, but are the necessary items in determining if the transition to the Global Analysis phase can be made.

Several benefits are visible in the proposed approach for Local Analysis and these are outlined below.

- *Incremental and iterative development*: Provides a better focus on distinct functions because they are analyzed one at a time. In addition, it is also easier to measure progress and validate requirements piecewise.

- *Early V&V*: Facilitates early detection and correction of requirements errors.

- *Nearness in time*: Enhances the stakeholders' recall capability, focus and feedback.

- *Better quality requirements*: The early V&V and the "nearness in time" factor enable the generation of better quality requirements early in the requirements phase.

- *Cost effective*: Because errors are detected and corrected early, there is a minimal propagation of errors to the later phases; this positively affects the cost and schedule constraints [34].

- *Well-defined*: The activities of the Local Analysis phase have been selected such that they have clear objectives and provide the seamless evolutionary path for requirements. We have also included two critical analysis activities – Modeling, and Rationalization and Justification – that are often overlooked by current approaches. In addition, we introduce the concept of local conflict resolution to address inconsistencies and incompatibilities in individual sets of requirements.

## 2.3 Global Analysis Phase

The Global Analysis phase follows the Local Analysis phase and it forms the second component of the expanded RGM. The objectives of the activities defined in the Global Analysis phase dictate that the elicited requirements be examined as a single comprehensive set rather than as individual subsets. This phase includes two sub-components (illustrated in Figure 4):

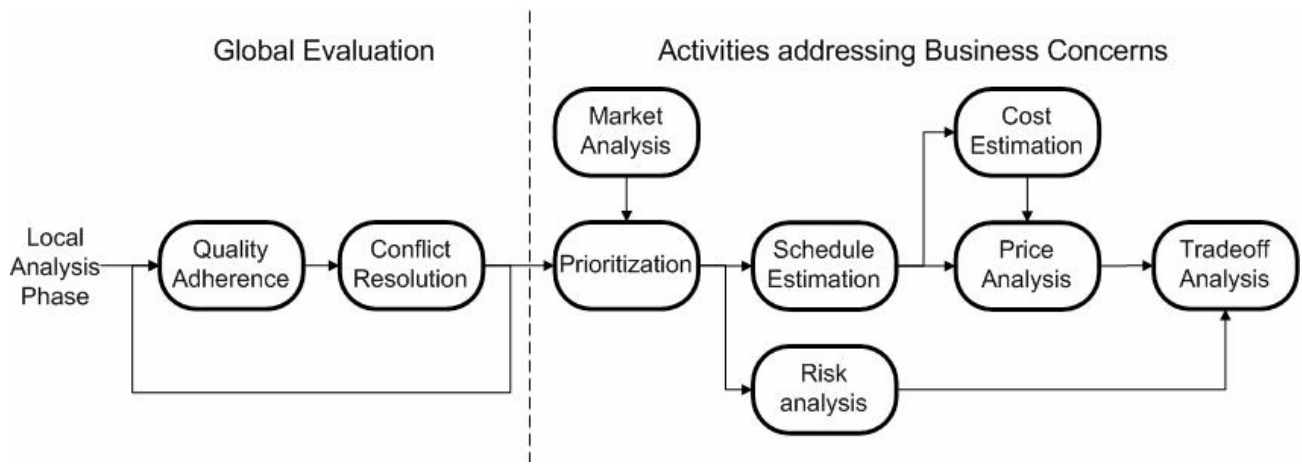

**Figure 4. Global Analysis phase and its activities**

- **Global Evaluation:** completes the verification process and resolves outstanding conflicts, and

- **Address Business Concerns:** assists in the evaluation of requirements from the business perspective.

## 2.3.1 Global Evaluation Component

As seen from Figure 4, the first two activities are concerned with a holistic evaluation of the requirements, and are iterative in nature. Because verification during Local Analysis cannot completely address certain quality characteristics (completeness, traceability and inconsistency), we have introduced the Quality Adherence activity during Global Analysis to complement the requirements verification conducted in the preceding Local Analysis phase. Traceability checking, which is included in the Quality Adherence activity, is simplified by the requirements rationale document produced by the prior Rationalization and Justification activity. Coupled with the fact that the Quality Adherence activity concentrates primarily on the linkages between sets of related requirements, with minimal focus on requirements *within* the sets, the effort required to complete the verification process is substantially reduced. Moreover, the concentration on the linkages improves focus and enhances detection of requirements errors. Inconsistencies identified during verification are addressed in the Conflict Resolution activity in a non-confrontationist atmosphere.

The strengths of the Global Evaluation component of the Global Analysis phase are:

- *Earlier verification:* The Global Evaluation component verifies the complete set of requirements as soon as they emerge from the Local Analysis phase and thus, positively affects the cost and quality of requirements.

- *Reduced effort and more focused scope*: These benefits are a result of the Global Evaluation component focusing primarily on links among sets of related requirements.

- *Optional validation*: Although requirements validation is not an objective of the Global Evaluation process, if the customer so desires, validation can be performed to substantiate that the proposed system does reflect its intended purpose.

## 2.3.2  Global Concerns Component

Business concerns are critical in identifying the final set of requirements for any system. Hence, our model focuses on these concerns in the latter half of the Global Analysis phase. The Business Concerns component helps determine project feasibility and scope, and helps address organizational issues and constraints. The defined activities are an extension to the Requirements Triage model proposed by Alan Davis [4]. However, unlike the Triage model which has a narrower focus, our approach covers the entire requirements process and provides for a seamless integration of the Business Concerns component with the rest of the model.

A brief description of the activities in this component is provided below:

a) **Market Analysis:** helps collect market information such as user expectation, market trends, competitor's product features, and so forth.

b) **Prioritization:** assists in ranking requirements based on importance to the user and the value added to the product. Priorities assigned can be categories (low, medium, high) or relative to one another [35]

c) **Risk Analysis:** focuses on examining the complete set of requirements for risk factors pertaining to product engineering, development environment and program constraints [36]. During this activity, the risk exposure$^2$ of the requirements is determined.

d) **Schedule Estimation**: assists in determining the development time of the components and identifying the critical components of the software system. PERT [37] and CPM [38] are most commonly used for schedule estimation.

e) **Cost Estimation**: helps predict the amount of work or effort required in developing the system. The size of the software is a major factor in determining the cost of the project. Size can be represented either in terms of lines of code or function points.

f) **Price Analysis**: focuses on deciding a fair and reasonable price for the product independent of the cost of individual components and proposed profit. Additionally, this activity also determines which functional capabilities are optional and can be dropped without affecting the product's value.

g) **Tradeoff Analysis**: helps evaluate the pros and cons of the system in Operational, Technical, Schedule, Economic, and Legal terms. In addition, tradeoff analysis also includes conflict resolution for requirements that are incompatible with the customer's constraints, e.g., cost and schedule.

---

$^2$ Risk exposure is the product of the likelihood that the risk will occur and the magnitude of the consequences of its occurrence.

Thus, after completion of the second half of the Global Analysis phase, we obtain conflict-free set of requirements that meet customer intent and which have well-defined scope. Moreover, the business concerns component supports better management through the generation of timely and appropriately scoped information.

The benefits of the Business Concerns component are:

- *Better focus*: Helps maintain a focused analysis on information related to organizational and management constraints.

- *Early detection of conflicts*: The identification of requirement incompatibilities relative to the business constraints is performed earlier in the requirements phase, that is, it is not postponed until the generation of the formal SRS.

- *Facilitates better management decisions*: Because the quality of the requirements has been established prior to these activities, generated estimates will tend to have less deviation from their actual counterparts. Thus, management can make better informed decisions, and thereby increase the probability of project success.

- *Distinct conflict resolution objectives*: Our model establishes a clear distinction among the conflict resolution activities in the Global Analysis and Local Analysis phases. During Local Analysis and Global Evaluation, conflicts in requirements stem from inconsistencies. However, in the latter half of the Global Analysis phase, conflict resolution involves negotiating incompatibilities between requirements and customer constraints. In explicitly recognizing this difference, we are able to apply a more appropriate set of conflict resolution techniques.

Although we have not yet experimentally validated the enunciated benefits stemming from our proposed Local and Global Analysis approaches, they can be substantiated through literature citations and rationalization based on experiences in the industry. Ackerman *et.al* and Russel have shown a substantial net benefit from formal inspections (verification) of *intermediate* work products in software development (as well as for systems in general) [39] [40]. Boehm *et.al.* and Jeffery also state that early validation and correction of user requirements can alleviate many of the problems associated with software development, particularly during the maintenance phase [14][15]. In addition, it is well acknowledged that it is much cheaper to detect and fix these errors early in the software development life cycle than later [34]. This is supported by Boehm, who shows that the cost of correcting errors in later phases of the development lifecycle escalates nearly 200 times [41].The above research clearly points in the direction that early verification and validation results in the earlier correction of errors. This in turn, reduces the cost incurred in fixing those errors. Our model applies this proven principle to the requirements phase by performing V&V on *incremental sets of requirements*. That is. it enables us to detect errors much earlier in the requirements phase as compared to performing conventional V&V towards the end after the generation of the SRS.

Moreover, the benefits of clarity of information and better feedback attributed to the "nearness in time" factor are supported by research in the psychology field. Ebbinghaus shows that the rate of forgetting is logarithmic - this is popularly known as the "forgetting curve" [42]. Jenkins and Dallenbach conducted a study which shows that interference causes forgetfulness [43]. Since interference accumulates over time, the retention capability of a person also reduces over time. The decay in memory retention is also confirmed by the experiments conducted by Bahrick [44]. Substantial evidence that supports the theory of increased forgetfulness with time can be found in psychology literature [45][46][47].

From the discussion in the previous two sections, it is evident that our model enhances the quality of generated requirements in a cost-effective manner. A more detailed description of the models can be found at [48].

# 3. OBJECTIVE-DRIVEN METHOD IDENTIFICATION

Our second research goal in refining the RGM was to simplify the task of selecting an appropriate set of methods for the requirement generation process by mapping methods to the activities in the expanded RGM. This mapping is achieved based on the activity objectives. In, Section 3.1 we explain the process of identifying the activity objectives. Section 3.2 illustrates the synchronization of methods and activity objectives.

## 3.1 Identification of Activity Objectives

In order to synchronize methods/techniques to requirements engineering activities, it is crucial to determine the objectives of each activity; this is because activity objectives drive the selection of methods. Hence, one of our primary concerns during the decomposition process was to identify the more fundamental, underlying activities and explicitly enunciate their objectives. Most of the current models such as RE Process Framework [5], and Win-Win model [3] provide a high-level perspective of the requirements generation process. Although higher-level activity objectives may be explicitly stated in these models, lower-level activities and objectives are often implied or ignored. For example, modeling, and identifying the rationale are often implied objectives of requirements analysis. Thus, within the expanded RGM model we have attempted to define activities at an adequate level of abstraction supporting the clear, explicit enunciation objectives. The activities in our expanded model are characterized by the attributes described in the following table [49]:

| **Attribute** | **Description** |
| --- | --- |
| Activity Name | Name of the activity |
| Objective | Goal/aim of the activity |
| Action Points | Milestones to be achieved by the activity |
| Pre-condition | Conditions to be satisfied before activity commencement |
| Doer | Person conducting the activity |
| Participants | Participants in the activity |
| Input documents | Documents needed for the activity to begin |
| Output documents | Documents produced at the completion of the activity |

**Table 1. Activity attributes**

Thus, each activity in the expanded RGM is well-defined, having enunciated objectives that facilitate the mapping of methods to activities based on those objectives. An example of an activity, its objective and other attributes is illustrated below (Table 2):

| Activity Name | Rationalization & Justification |
|---|---|
| Objective | Find rationale, justify, refine and decompose requirements |
| Action Points | Identify classification of requirements (functional/ non-functional) |
| | Address the question "why" underlying the requirements |
| | Identify requirements which are high level or unspecific |
| | Justify the requirements |
| | Refine and decompose the high level requirements |
| Pre-condition | Elicitation activity completed |
| Doer | Requirements Engineer |
| Participant | User, customer, developer |
| Resource/Input docs | Unstructured requirements, Domain Info, Organizational standards and regulations |
| Effect/Output docs | Non-prioritized requirement list |

**Table 2. Example of an activity and its attributes**

## 3.2 Mapping of Methods to Activity Objectives

Once the activity objectives are defined, the next step is to identify methods that help achieve the stated objectives of an activity. Our initial goal was to identify all possible techniques for the various activities in the model. We decided, however, to focus on methods commonly used in the industry. This is because the literature includes a large number of methods and only a fraction of those are actually being employed in the industry. Additionally, the goal of our research is to provide guidance to the requirements engineer in the "real world". Table 3 provides an illustration of our mapping of two activities (Requirements Elicitation Meeting and Risk Analysis) to their respective objectives and subsequently, to the applicable methods supporting the achievement of those objectives.

| **Activity Name** | **Activity Objective** | **Applicable Methods** |
|---|---|---|
| Requirements Elicitation Meeting | Correctly identify and capture requirements from the stakeholders | Interviews, Observation, Task Demonstration, Document Studies, Questionnaires, Brainstorming, Focus Groups, Requirements Workshops, Prototyping [10]. |
| Risk Analysis | Estimate risk in the development of system components | Criticality Analysis [50], Fault Tree Analysis [51], Risk Reduction Leverage [52], Event Tree Analysis [53], Monte Carlo Simulation [54], FMECA (Failure mode, effects, and criticality analysis) [55]. |

**Table 3. Mapping of methods to activities based on their objectives**

The complete mapping of methods to all activities in our proposed model is provided in [48]. A total of 77 methods have been identified for the 14 activities in the expanded RGM. As a result of this mapping we achieve two goals. First, we have synchronized effective methods with well-defined and appropriately decomposed activities. In comparison, the requirements engineering literature identifies a large number of methods for high-level activities with no clear mapping between methods and objectives. Secondly, we provide a reduced set of methods for the activities and thus, make the task of selecting appropriate methods for an activity easier.

In addition to identifying the methods, we have also analyzed these methods and provided a detailed description of them. The methods are evaluated relative to their pros and cons to facilitate the selection process. Furthermore, the process of implementing each method is also explained in detail. A tabulated version of one of the methods is shown in Table 4. Detailed explanations of all the methods can also be obtained at [48].

| **Critical Path Method** ||
|---|---|
| provides a means for determining [56] <br> (1) which jobs or activities, are "critical" in their effect on total project time, and <br> (2) how best to schedule all jobs in order to meet a target date at minimum cost ||
| Pros | • CPM chart is an excellent tool for communication <br> • Shows dependencies of different aspects of the project <br> • Focuses attention on critical aspects of the plan <br> • Employing CPM enables management to use resources more wisely <br> • Greatly improved control over complex development work and production programs <br> • A good scheduling tool for large and small projects |
| Cons | • CPM's single time estimate fails to consider the effects of variability in path completion times <br> • Not adequate for expressing all of the various sophisticated relationships between nodes <br> • Expensive method <br> • Time consuming |

**Table 4. Snapshot of one of the identified methods**

## 3.3 Advantages of Method Synchronization

Several benefits are apparent as a consequence of mapping methods to activities; they are outlined below:

1) **Explicitly stated objectives:** Our decomposition process identifies activities that have clear, focused objectives. Every activity objective is stated explicitly - not implicitly, which is prevalent in many of the current models because of their high level focus on the requirements generation process.

2) **Synchronization of methods with activities:** We have mapped the commonly used methods in the industry to the activities based on activity objectives. Additionally, unlike previous method mapping research, our work is based on an appropriate level of activity decomposition illustrating a clear synchronization between methods and activity objectives.

3) **Easier method selection:** Because we provide a smaller set of methods for each activity, it is easier for the requirements engineer to select an appropriate set of overall methods.

Even though we have identified a reduced set of methods for the requirements generation process, the number of methods for each activity is still substantial. From Table 3, we can see that there are nine methods to choose from for the Elicitation activity and six for the Risk Analysis activity. In a real world scenario, the requirements engineer makes his/her choice based on certain method selection criteria and by weighing the strengths and weakness of each method against that criterion. Our third goal, as detailed in the next section, is to simplify this task.

# 4. CRITERIA BASED METHOD SELECTION

A substantial part of our research has been to identify an appropriate set of methods that support individual requirements engineering activities and the achievement of their stated objectives. Selecting the one method within that set, however, is often based on operational or organizational criteria that have little bearing on the activity objectives. For example, we might desire to select the method that has the least cost to implement. Hence, to provide guidance in the method selection process, we have chosen four criteria commonly used in the industry, and have analyzed each method relative to its ability to support the achievement of that criteria. The four criteria are introduced next, followed by method analyses and a selection process.

- **Personnel**: Selection based on the number of people involved and their expertise. This criterion is the most widely used criterion in the selection of methods because a project usually has limited work staff.

- **Time**: Selection based on the time needed to perform a technique. Often the requirements engineer has very limited time to perform a particular activity. In such situations, it is necessary to employ the technique that achieves the activity objective in the least amount of time.

- **Cost**: Selection based on expenses involved in conducting the method. In situations where the project is under budgeted, it is imperative to select techniques that minimize the cost incurred.

- **Completeness**: Selection based on the coverage of activity objectives. This criterion is used when it is necessary to completely achieve the objective of one activity before proceeding to the next activity, e.g. life critical systems.

For each of the above criteria, we identified the methods for each activity in the expanded RGM that best optimize each selection criteria. This results in smaller, categorized sets of methods for each activity, which in turn simplifies the selection task of the requirements engineer. The concept of using the selection criteria in choosing methods for an activity is illustrated in Figure 5. The bubbles C1, C2, C3 and C4 represent the set of methods which satisfy each of the four selection criteria. For example, C1 may represent methods optimizing cost criteria, C2 – methods satisfying completeness criterion, and so forth. The three activities depict three different scenarios in which the set of methods accommodates the four selection criteria:

- **Ideal scenario:** This occurs when there are a set of methods which satisfy all the four criteria as shown by the intersection of the four bubbles in Activity 1. In this situation, the requirements engineer's task of method selection is the easiest with the selected method(s) satisfying all criteria.

- **Normal scenario:** Activity 2 illustrates the normal scenario where there exist a set of methods which satisfy some criteria but not all. As seen from Activity 2, C4 intersects with C1 and C3, but not with C2. Also, C2 does not intersect with any

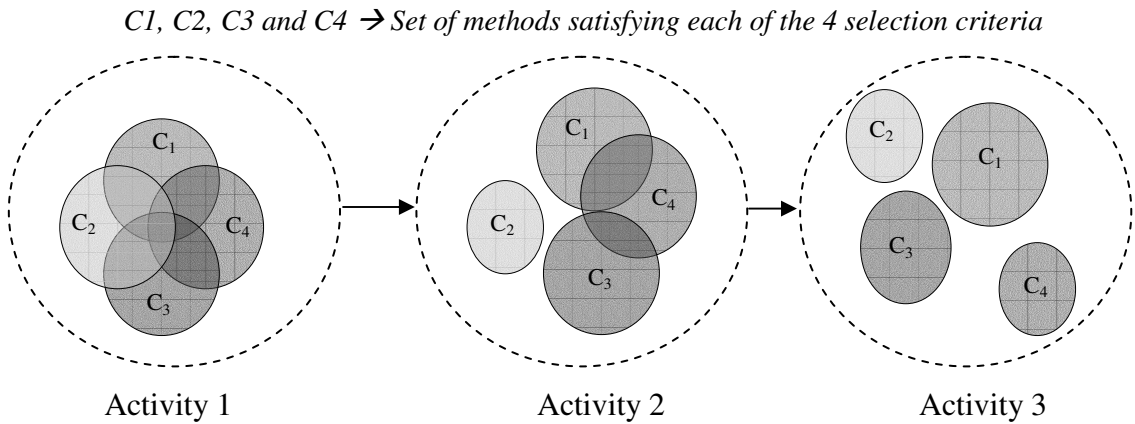

*C1, C2, C3 and C4* → *Set of methods satisfying each of the 4 selection criteria*

Activity 1          Activity 2          Activity 3

**Figure 5: Use of selection criteria in choosing methods**

of the other bubbles. Hence, in such situations, the choices of methods often satisfy some criteria but compromise on the others.

- **Worst case scenario:** This occurs when the methods satisfy different criteria as shown in Activity 3. Consequently, the requirements engineer must select methods that satisfy only one criterion while compromising the rest.

Thus, given the framework where the methods supporting a single activity are grouped based on selection criteria, the task of the requirements engineer is simplified because s/he can select from a reduced set of methods that optimize the selection criteria. Suppose the requirements engineer needs to identify methods that optimize the cost criterion. Suppose in Figure 6, the bubble C1 represents methods satisfying cost criterion, then the requirements engineer has to only examine those methods in bubble C1 for each of the three activities in order to select the appropriate methods. Additionally, within the set of methods satisfying the chosen criterion, the methods can be selected by comparing them based on their documented strengths and weaknesses. The arrows in the figure illustrate this selection process. Furthermore, this framework and approach enables the requirements engineer to achieve better results by facilitating the selection of methods that satisfy multiple criteria. Another advantage is that the number of methods employed

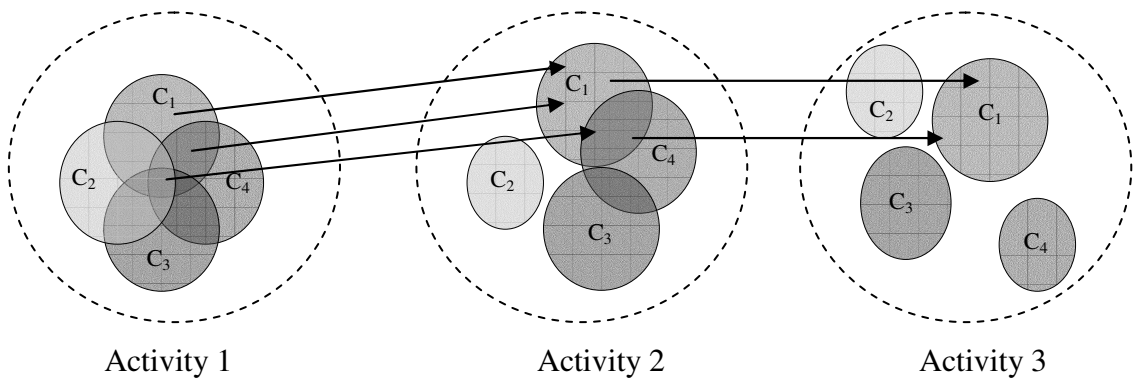

Activity 1　　　　　　　　　　Activity 2　　　　　　　　　　Activity 3

**Figure 6: Possible paths of methods based on one selection criterion**

can be minimized by selecting a minimal set of methods that meet the objectives of several activities. The following example illustrates the utility and flexibility of the selection process outlined above.

## 4.1 An Example

In this section we illustrate the selection process and the usefulness of the method selection criteria. Here we consider the Risk Analysis activity which assists in determining the risk involved in the development of the software components. Based on the literature and industry practice, we identified six risk analysis methods that are widely used in software project development. These methods were then analyzed to determine those which best optimize cost, time, personnel, and completeness. The mapping of methods to the Risk Analysis activity based on the four selection criteria is depicted in Table 5.

| Risk Analysis Activity | |
|---|---|
| **Selection Criteria** | **Methods for the Risk Analysis activity** |
| Personnel | FMECA (Failure mode, effects, and criticality analysis), Monte Carlo Simulation |
| Time | Criticality Analysis, Monte Carlo Simulation |
| Cost | FMECA (Failure mode, effects, and criticality analysis), Criticality Analysis |
| Completeness | Monte Carlo Simulation, Fault Tree Analysis and Event Tree Analysis |

**Table 5: Methods optimizing selection criteria for the Risk Analysis activity**

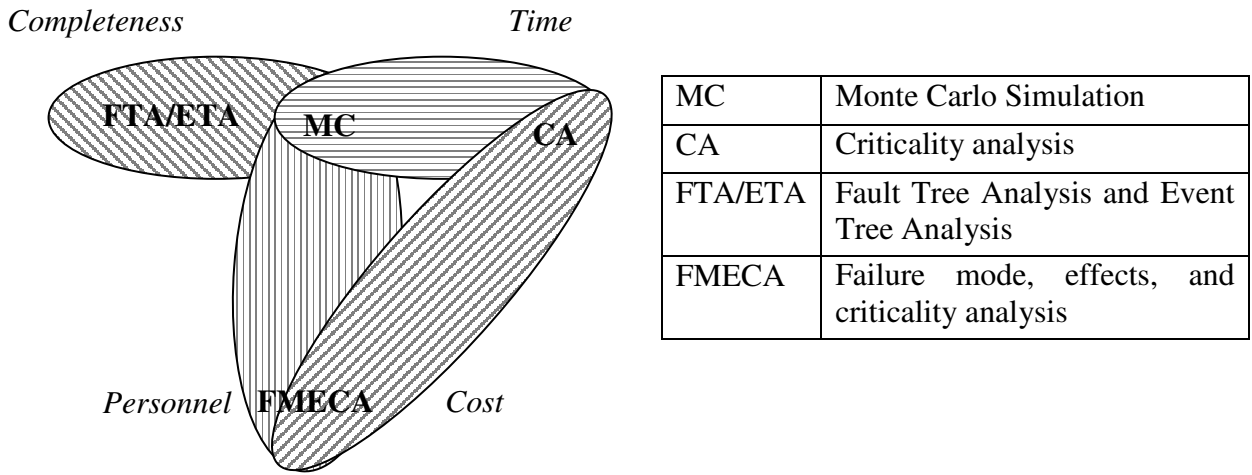

**Figure 7: Methods optimizing multiple criteria for the Risk Analysis activity**

| Criteria | Methods for Risk Analysis Activity | | | |
|---|---|---|---|---|
| | **FMECA** | **MC** | **CA** | **FTA & ETA** |
| **Personnel** | X | X | | |
| **Cost** | X | | X | |
| **Completeness** | | X | | X |

**Table 6: Methods optimizing multiple criteria for the Risk Analysis activity**

As seen from Figure 7 and Table 6, Monte Carlo Simulation technique optimizes three selection criteria – time, personnel and completeness. Hence, this technique is the preferred method because it optimizes three selection criteria. However, this does not imply that Monte Carlo Simulation is the best for all the three criteria taken separately. In fact, if time is the deciding factor and other criteria are marginalized, Criticality Analysis is a better choice than Monte Carlo Simulation. This can be judged by analyzing the pros and cons of both of these methods. The method, Criticality Analysis, optimizes both the time and cost criteria. In situations where there is a primary and secondary criterion, such as time and cost, Criticality Analysis is the most suitable method.

Another consideration in the selection of methods is to employ a *minimal* set of methods for the requirements generation process. Our framework facilitates this goal by identifying methods that optimize the selection criteria for all activities in the requirements process and thus, enables the requirements engineer to make informed decisions. For example, if interview is the preferred method for the Elicitation activity and is the second preference for the Rationalization and Justification activity, the requirements engineer can select interviews for the latter activity to reduce the overhead of performing two different methods. Thus, from the discussed example, we see that the method selection task of the requirements engineer is simplified through two features – (1) documentation of the pros and cons of the methods and, (2) the grouping of methods based on selection criteria.

Our framework facilitates the selection of a path of methods that optimize the chosen selection criteria for the entire requirements generation process. Using Completeness as a criterion Table 7 illustrates a mapping of methods to activities in the Business Concerns component, starting with the Risk Analysis activity.

| Completeness Criteria Applied to Activities of the Business Concerns Component | | | | |
|---|---|---|---|---|
| **Risk Analysis** | **Cost Estimation** | **Schedule Estimation** | **Price Analysis** | **Tradeoff Analysis** |
| Monte Carlo Simulation<br><br>Criticality Analysis | COCOMO II [57]<br><br>Function Point Approach [58] | PERT (Program Evaluation and Review Technique)<br><br>CPM (Critical Path Method) | Comparative Price Analysis [59]<br><br>Value Analysis [60] | PMI (Plus, Minus, and Implications) [61]<br><br>Decision Analysis [62]<br><br>Internal Rate of Return [63]<br><br>Net Present Value [64] |

**Table 7: Mapping of methods based on completeness criterion**

As depicted, there are a small set of methods for each activity. The choice among these methods can be made by studying the documented strengths and weaknesses of the methods. This approach certainly has an advantage over the current situation where the requirements engineer is provided with a large collection of methods mapped to high level activities having implicit objectives. Based on completeness criterion, a sample path of methods for the activities in Table 6 could be:

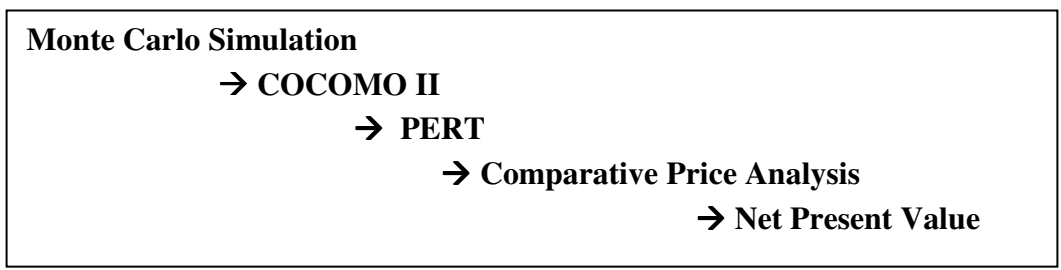
**Monte Carlo Simulation**
  → **COCOMO II**
    → **PERT**
      → **Comparative Price Analysis**
        → **Net Present Value**

Such a table is provided for each of the four criteria and facilitates the making of informed decisions on method selection for a particular activity. By selecting a specific criterion, our research has shown that for the requirements generation process we can remove from consideration nearly one-half of the 77 potential methods. For each activity this then translates into a reduced set of typically 2-3 methods from which to choose. Thus, our research enables the requirements engineer to focus on selecting the most appropriate method that not only satisfies the project constraints but also the activity objectives.

Thus, our proposed framework provides guidance in the selection of methods by – prescribing a reduced set of effective methods that optimize the selection criteria, and facilitating selection among these methods through the documentation of their pros and cons.

## 5. CONCLUSION

Our research has been guided by three primary objectives – (1) to propose a well-defined model having documented activities reflecting adequate levels of decomposition (2) to identify principal activity objectives and map popular methods to those activities according to their objectives, and (3) to enhance the guidance in selecting methods for activities in the requirements generation process. We propose a model that addresses the complete requirements generation process, and which identifies activities at an appropriate level of activity abstraction. The decomposition in our model illustrates a clear evolutionary path of the requirements. In addition, it also facilitates the mapping of methods to activities. In addition to overcoming the limitations of current models, our proposed model supports the generation of quality requirements in a cost-effective manner. This is achieved by performing the V&V activities iteratively on smaller sets of requirements as they relate to individual system functionalities.

We also identify the objectives of the activities and explicitly state them. The identification of objectives is imperative because the mapping of methods to activities is driven by the objectives. We have also identified methods that are commonly used in the industry and have mapped them to activities based on activity objectives. Thus, we achieve a synchronization of methods and activity objectives, unlike current research that only maps methods to high-level activity objectives. Additionally, we have analyzed the methods and have provided a detailed description of the technique, their benefits and drawbacks. Subsequently, for each activity the requirements engineer now has a more appropriate set of methods from which s/he can select.

To enhance the guidance in method selection, we have identified four selection criteria (cost, time, personnel, and completeness) that are widely used in the industry and have grouped the methods for each activity based on these criteria. As a result, for each activity we have identified a smaller set of methods from which to choose that optimize each of the four selection criteria. This setting enhances the guidance in selecting the most appropriate method for an activity based on a selection criterion or a combination of criteria. In effect, we have provided a framework that enables the requirements engineer to make informed decisions during the method selection process.

The benefits attributed to our model are substantiated through literature citation and rationalization based on experiences in the industry [14][15][40]. Our claims are further strengthened by the fact that our model extends the Requirements Generation Model (RGM) [16], whose effectiveness has been empirically validated. In the future, we envision a detailed empirical evaluation to provide better insights into the implementation aspects of the expanded RGM model and our approach. In addition, continuing the mapping of methods for the remainder of the software development life cycle can provide additional guidance for the software engineering community, and subsequently improve the success rate of software projects.

[55]  Military Standard, "*Procedures for Performing a Failure Mode, Effects and Criticality Analysis*", MIL-STD-1629A, 12 June 1977.

[56]  Hulett, David T. (1995). "*Project schedule risk assessment*", Project Management Quarterly, 26 (1), March, 21-31.

[57]  Nancy Merlo, Schett, 2002, "*COCOMO (Constructive Cost Model)*", Seminar on Software Cost Estimation WS 2002 / 2003, 2002.

[58]  Roger Heller, "*An Introduction to Function Point Analysis*", http://www.qpmg.com/fp-intro.htm, 2002.

[59]  The Federal Aviation Administration Acquisition System Toolset, http://fast.faa.gov/pricing/98-30-C5.htm, 2004.

[60]  USaid, "*Primer and Checklist for Conducting Cost and Price Analysis for Interagency Agreements*", http://www.usaid.gov/policy/ads/300/306maa.pdf, October 2002.

[61]  Edward de Bono, "*Serious Creativity*", HarperBusiness, New York, US, 1992.

[62]  R.T. Clemen, "*Making Hard Decision: An Introduction to Decision Analysis*", 2nd Edition, Belfont, California, Duxbury Press, 1996.

[63]  Ignacio Velez-Pareja, "*The Weighted Internal Rate of Return (WIRR) and Expanded Benefit-Cost Ratio (EB/CR)*" (September 19, 2000). Universidad Javeriana Working Paper No. 9. http://ssrn.com/abstract=242867

[64]  Duncan Williamson, "*Capital Budgeting: The Key Numerical techniques*", http://www.duncanwil.co.uk/invapp.html, October 2001.

# 7. FIGURES

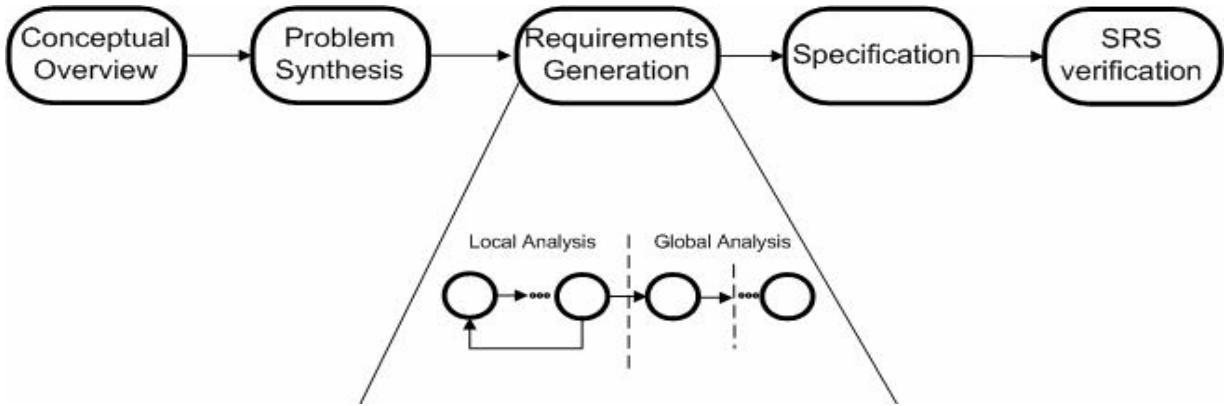

**Figure 1. The Local and Global Analysis phases in the context of the RGM**

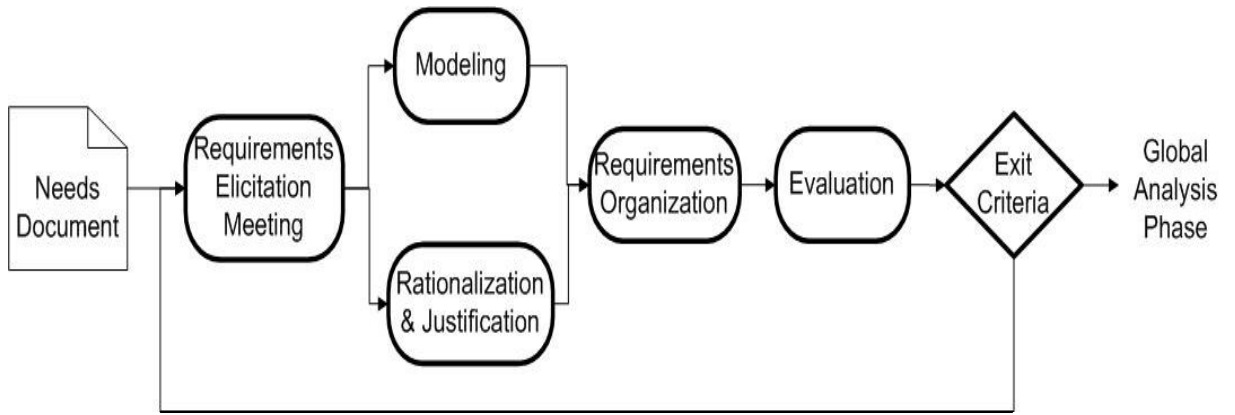

**Figure 2. Local Analysis phase and its activities**

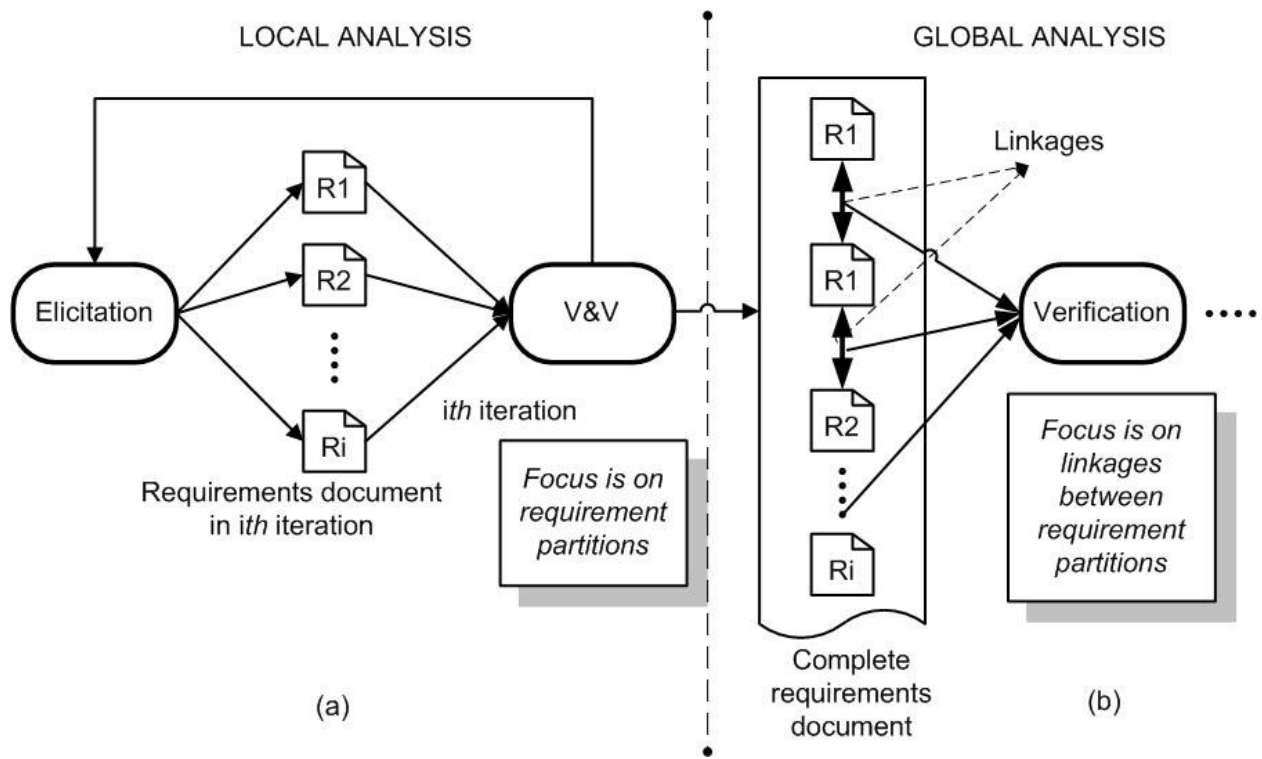

**Figure 3. V&V focus in Local and Global Analysis phases**

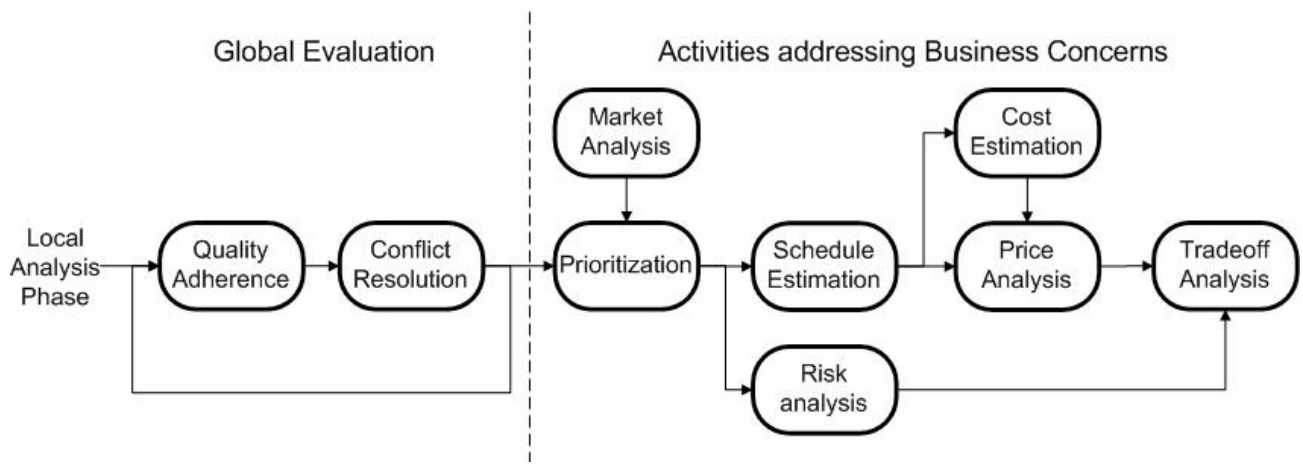

**Figure 4. Global Analysis phase and its activities**

*C1, C2, C3 and C4 → Set of methods satisfying each of the 4 selection criteria*

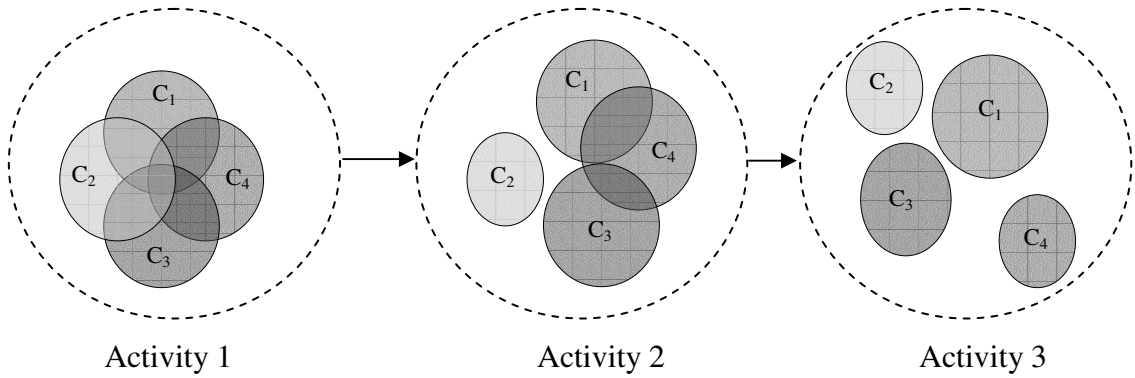

**Figure 5: Use of selection criteria in choosing methods**

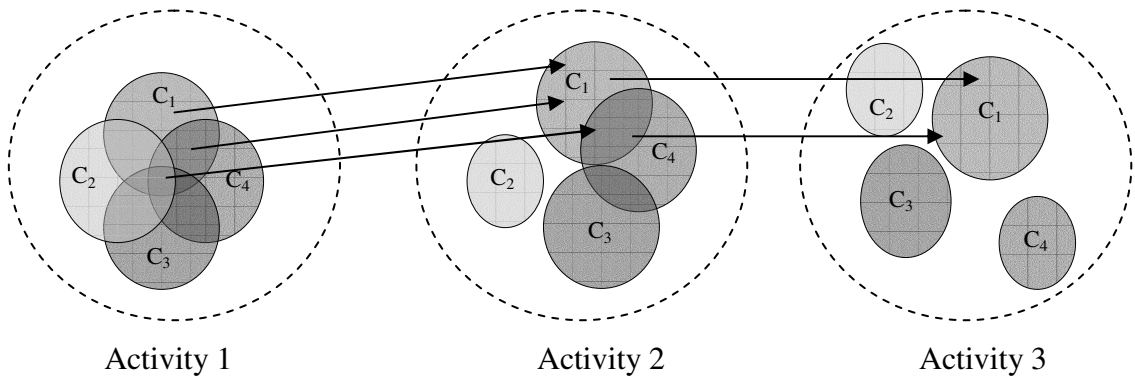

**Figure 6: Possible paths of methods based on one selection criterion**

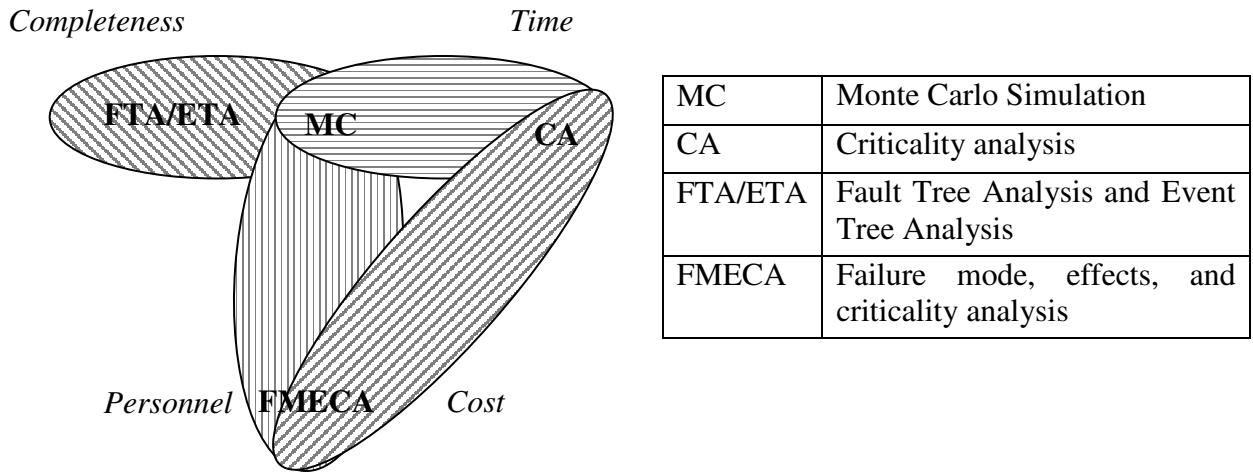

**Figure 7: Methods optimizing multiple criteria for the Risk Analysis activity**

# 8. LIST OF FIGURES